\documentclass[12pt]{article}
\usepackage{amsmath}
\usepackage{epsf}
\usepackage{epsfig}
\usepackage{here}
\usepackage{amssymb}
\usepackage{citesort}
\usepackage{graphicx}
\usepackage{latexsym}
\newcommand{\be}{\begin{equation}}
\newcommand{\ee}{\end{equation}}
\newcommand{\bear}{\begin{eqnarray}}
\newcommand{\ear}{\end{eqnarray}}

\newsavebox{\LSIM}
\sbox{\LSIM}{\raisebox{-1ex}{$\ \stackrel{\textstyle<}{\sim}\ $}}
\newcommand{\lsim}{\usebox{\LSIM}}
\newsavebox{\GSIM}
\sbox{\GSIM}{\raisebox{-1ex}{$\ \stackrel{\textstyle>}{\sim}\ $}}
\newcommand{\gsim}{\usebox{\GSIM}}

\begin{document}
\begin{titlepage}
\begin{flushright}
BA-00-42\\
hep-ph/0010195
\end{flushright}
$\mbox{ }$
\vspace{.1cm}
\begin{center}
\vspace{.5cm}
{\bf\Large Fermion Masses, Mixings and Proton}\\[.3cm]
{\bf\Large Decay in a Randall-Sundrum Model}\\
\vspace{1cm}
Stephan J. Huber\footnote{shuber@bartol.udel.edu} 
and
Qaisar Shafi\footnote{shafi@bartol.udel.edu} \\

\vspace{1cm} {\em 
Bartol Research Institute, University of Delaware, Newark, DE 19716, USA
}
\end{center}
\bigskip\noindent
\vspace{1.cm}
\begin{abstract}
We consider a Randall-Sundrum model in which the Standard Model 
fermions and gauge bosons correspond to bulk fields. We show how the
observed charged fermion masses and CKM mixings can be explained, 
without introducing hierarchical Yukawa couplings. We then
study the impact on the mass scales associated with non-renormalizable
operators responsible for proton decay, neutrino masses, and flavor changing
neutral currents. Although mass scales as high as $10^{11}- 10^{12}$ GeV 
are in principle possible, dimensionless couplings of order $10^{-8}$   are
still needed to adequately suppress proton decay.
Large neutrino mixings seem to require new physics
beyond the Standard Model.
\end{abstract}
\end{titlepage}
\section{Introduction}
Higher dimensional models of space-time with non-factorizable geometries 
have attracted much interest recently, especially because they may provide a 
solution to the gauge hierarchy problem.  In the Randall-Sundrum approach 
\cite{RS} (see also \cite{G}) the warp factor $\Omega=e^{-\pi k R}$
generates an exponential hierarchy between the
effective fundamental mass scales on the two branes located 
at the orbifold fixed points in the extra dimension. If the brane separation
is $kR\simeq 11$, the scale on one brane is of TeV-size,
while the scale on the other brane is of order 
$M_{\rm Pl}\sim 10^{19}$ GeV.  The AdS curvature $k$ and the 5d Planck
mass $M_5$ are both of order $M_{\rm Pl}$.

In the original proposal \cite{RS} the SM fields are all
assumed to reside on the TeV-brane and  only gravity propagates
in the extra dimension. This setup is very economical in 
solving the hierarchy problem. However, one would naively expect
that non-renormalizable operators in the 4d effective theory, 
now only  TeV-scale suppressed, would lead to rapid proton decay,
and unacceptably large neutrino masses and flavor changing neutral 
currents.  

In the following we will explore to what extent this problem can
be solved by moving the SM fermions away from the TeV-bane,
without invoking ad hoc symmetries such as baryon and/or lepton
number. The basic idea, already explored in ref.~\cite{GP},
exploits the fact that closer to the Planck-brane the effective fundamental 
scale is much larger than a TeV. However, the fermions cannot be moved  
arbitrarily far from the TeV-brane, 
in case their overlap with the Higgs field becomes too small
to provide sufficiently large fermion masses. The Higgs field
must reside near the TeV-brane if the model is expected to solve the 
hierarchy problem \cite{CHNOY,HS,DHR} (supersymmetry may help 
to avoid this constraint). The SM gauge bosons are assumed to be 
bulk fields to maintain gauge invariance.
Taking into account constraints from the electroweak precision data 
\cite{GP,CHNOY,HS,DHR,DHR2,DHRP,dAS}, 
especially the W and Z boson mass ratio \cite{HS}, the Kaluza Klein (KK) 
excitations of bulk gauge bosons and fermions are of order 10 TeV.
Thus, some tuning of parameters is required to reproduce the measured
W and Z boson masses. 

Having the SM fermions in the bulk can also help explain the fermion mass
hierarchy \cite{GP,AS,kt}. Heavy fermions are localized near the TeV-brane,
where their overlap with Higgs fields is large, while the light fermions
reside closer to the Planck-brane. We will demonstrate that this mechanism 
also generates acceptable quark mixings without invoking any flavor symmetry. 
The generation of fermion masses
and quark mixings along these lines has already been explored in the case of flat 
extra dimensions \cite{AS,kt}. There, some additional physics has to be
assumed in order to localize the fermions or provide an appropriate 
profile for the Higgs in the extra dimension. In the scenario considered
here, these are automatically generated by the non-factorizable geometry.  

In the following we present a set of ``order one parameters'', describing 
the localization of the bulk fermions, which successfully reproduces 
the measured fermion masses and quark mixings. We also 
study the impact on the mass scales associated with  non-renormalizable
operators responsible for proton decay, flavor changing neutral currents, 
and neutrino masses. Finally, we briefly discuss the issue of neutrino mixings.

\section{5d fermions}
We consider the non-factorizable metric \cite{RS}
\begin{equation} \label{met}
ds^2=e^{-2\sigma(y)}\eta_{\mu\nu}dx^{\mu}dx^{\nu}+dy^2,
\end{equation}
where $\sigma(y)=k|y|$.  
The 4-dimensional metric is $\eta_{\mu\nu}={\rm diag}(-1,1,1,1)$, 
$k$ is the AdS curvature, and $y$ denotes the fifth coordinate.
The equation of motion for a fermion in curved space-time reads
\begin{equation}
E_a^M\gamma^a(\partial_M+\omega_M)\Psi+m_{\Psi}\Psi=0,
\end{equation}
where $E_a^M$ is the f\"unfbein, $\gamma^a$ are the Dirac
matrices in flat space, and $\omega_M$ is the spin connection.
The index $M$ refers to objects in 5d curved space,
the index $a$ to those in tangent space. 
Under the $Z_2$ orbifold symmetry the fermions behave as
$\Psi(-y)_{\pm}=\pm \gamma_5 \Psi(y)_{\pm}$. Thus $\bar\Psi_{\pm}\Psi_{\pm}$ is odd 
and the Dirac mass term, which is also odd, can be parametrized 
as $m_{\Psi}=c\sigma'$. 
Here and in the following prime denotes differentiations with respect
to the fifth coordinate.
On the other hand, $\bar\Psi_{\pm}\Psi_{\mp}$ is even.
Using the metric (\ref{met}) one obtains for the left- and right-handed components
of the Dirac spinor \cite{GN,GP}
\begin{equation}
[e^{2\sigma}\partial_{\mu}\partial^{\mu}+\partial_5^2-\sigma'\partial_5-M^2]e^{-2\sigma}\Psi_{L,R}=0,
\end{equation} 
where $M^2=c(c\pm1)k^2\mp c\sigma''$ and  $\Psi_{L,R}=\pm\gamma_5\Psi_{L,R}$.

Decomposing the 5d fields as 
\begin{equation}
\Psi(x^{\mu},y)=\frac{1}{\sqrt{2\pi R}}\sum_{n=0}^{\infty}\Psi^{(n)}(x^{\mu})f_n(y),
\end{equation}
one ends up with the zero mode wave function \cite{GN,GP}
\begin{equation}
f_0=\frac{e^{(2-c)\sigma}}{N_0},
\end{equation}
where
\begin{equation}
N_0^2=\frac{e^{2\pi kR(1/2-c)}-1}{2\pi kR(1/2-c)}.
\end{equation} 
Because of the orbifold symmetry, the zero mode of 
$\Psi_+$ $(\Psi_-)$ is a left-handed (right-handed) Weyl spinor.  
For $c>1/2$ $(c<1/2)$ the fermion is localized near the boundary
at $y=0$ $(y=\pi R)$, i.e.~at the Planck-  (TeV-) brane.

\section{Masses for bulk quarks and leptons}
The zero modes of leptons and quarks acquire masses from their 
coupling to the Higgs field
\begin{equation} \label{3.1}
\int d^4x\int dy \sqrt{-g}\lambda^{(5)}_{ij}H \bar\Psi_{i+}\Psi_{j-}
\equiv \int d^4x ~ m_{ij} \bar\Psi_{iR}^{(0)}\Psi_{jL}^{(0)} +\cdots,
\end{equation}
where $\lambda^{(5)}_{ij}$ are the 5d Yukawa couplings. The 
4d Dirac masses are given by
\begin{equation} \label{3.2}
m_{ij}=\int_{-\pi R}^{\pi R}\frac{dy}{2\pi R}\lambda^{(5)}_{ij}e^{-4\sigma}H(y) f_{0i+}(y)f_{0j-}(y).
\end{equation}
We assume that the Higgs profile has an exponential form
which peaks at the TeV-brane
\begin{equation}
H(y)=H_0e^{ak(|y|-\pi R)}.
\end{equation}
Using the known mass of the W-boson we can fix one parameter,
which we take to be the amplitude $H_0$, in terms of the 
5d weak gauge coupling $g^{(5)}$. The parameter $a$
determines the width of the profile. If the profile satisfies the equations
of motion we have $a=4$ \cite{GW}, but we will also consider other values.  

Various constraints on the scenario with bulk gauge and fermion fields
have been discussed in the literature \cite{CHNOY,GP,HS,DHR,DHR2,dAS}.
With bulk gauge fields for instance, the SM relationship between the gauge couplings 
and masses of the Z and W bosons gets modified. The electroweak 
precision data then requires the lowest KK excitation of the gauge bosons 
to be heavier than about 10 TeV \cite{HS}.
This bound is independent of the localization of the fermions in the extra
dimension.  

Furthermore,  the KK excitations of the SM gauge bosons 
contribute to the electroweak precision observables. Their effect is 
relatively small if the SM fermions are localized towards the 
Planck-brane $(c>1/2)$. In this case KK masses around 
1 TeV are sufficient for the corrections from the gauge boson excitations 
to be compatible with the experimental bounds \cite{GP,DHR}.  Therefore, 
the above constraint derived from the W and Z boson mass ratio is very 
important in this range of parameters. In the following we will assume that
the mass of the first KK gauge boson is $m_1^{(G)}=10$ TeV.
The corresponding masses of the fermions are then in the range
10 to 13 TeV, for $0<c<1$.

Let us begin with the charged leptons. 
In the absence of Dirac masses for the neutrinos 
we can start with diagonal Yukawa couplings
for the leptons. To avoid a new hierarchy in the 
5d couplings, we assume $\lambda^{(5)}_{ii}=g^{(5)}$ 
(but will also discuss other possibilities below).
We take $k=M_5=\overline M_{\rm Pl}$, where $\overline M_{\rm Pl}=2.44\times10^{18}$
GeV is the reduced Planck mass. From $m_1^{(G)}=10$ TeV 
we determine the brane separation $kR=10.83$ \cite{DHRP},
and taking $a=4$ we obtain $H_0=0.396k/g^{(5)}$.

\begin{table}[t] \centering
\begin{tabular}{|c||c|c|c|} \hline
L& $e$ & $\mu$ & $\tau$  \\ \hline 
$m$ [MeV] & 0.5 & $106$ & $1777$  \\ \hline
(A):~$c_L$  & 0.681 & $0.591$ & $0.537$  \\ \hline
(B):~$c_L$  & 0.834 & $0.664$ & $0.567$  \\ \hline
(A):~$M_4$ [GeV] & $4.02~10^9$ & $1.98~10^7$ & $1.21~10^6$  \\ \hline 
\end{tabular} 
\caption{Lepton mass parameters in case of (A) left-right symmetry ($c_L=c_E$)
and (B) delocalized right-handed fermions ($c_E=1/2$).}
\label{t_L} 
\end{table}

\begin{figure}[t] 
\begin{picture}(100,160)
\put(95,-10){\epsfxsize7cm \epsffile{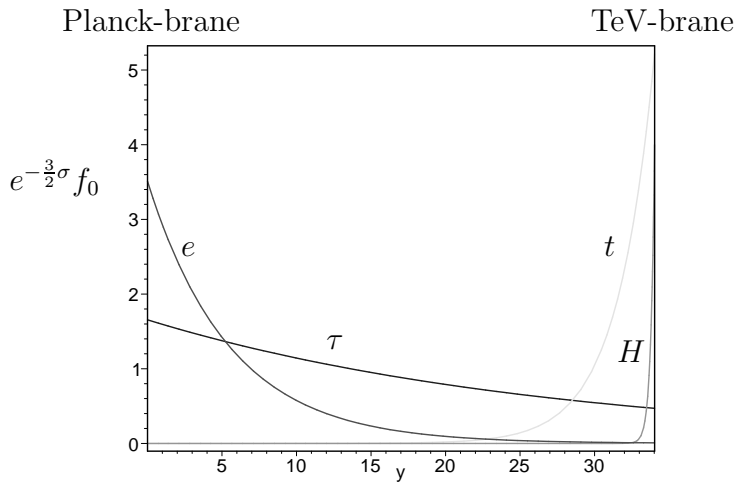}}
\put(50,100){{$e^{-\frac{3}{2}\sigma}f_0$}} 
\put(115,75){{$e$}} 
\put(170,40){{$\tau$}} 
\put(275,75){{$t$}} 
\put(280,35){{$H$}} 
\put(270,160){{TeV-brane}} 
\put(70,160){{Planck-brane}} 
\end{picture} 
\caption{Localization of the electron ($c=0.681$), tau ($c=0.537$)
and right-handed top quark ($c=0.1$) zero modes in the fifth dimension.
The Higgs profile $H$ is given for $a=4$ in units of $k^{3/2}$ and 
magnified by a factor of 10.}
\label{f_1}
\end{figure}

The lepton masses depend on the 5d mass parameters of the
left- and right-handed fermions, $c_L$ and $c_E$ respectively, 
which enter (\ref{3.2}).
The experimentally known lepton masses do not fix these six
$c$-parameters unambiguously. Hence we will concentrate on two 
special scenarios: (A) left-right symmetry, i.e.~$c_L=c_E$; (B) delocalized
right-handed fermions, i.e.~$c_E=1/2$. In table \ref{t_L} we give
the numerical values of the lepton mass parameters which reproduce
the physical masses in both cases. The larger the 5d Dirac mass, i.e.~$c$ 
parameter of the fermion becomes, the greater is its localization at the Planck-brane. 
Its overlap with the Higgs profile at the TeV-brane is consequently less, which 
is reflected in a smaller 4d fermion mass after electroweak symmetry breaking.
Our geometrical picture beautifully generates the charged 
lepton mass hierarchy by employing $c$-parameters of order unity.
In figure \ref{f_1} we present the wave functions of the 
electron and tau zero modes in the extra dimension, together with
the Higgs profile. One observes that the  electron, and to a lesser extent the tau, 
are  localized near the Planck-brane. The factor of $e^{-\frac{3}{2}\sigma}$ 
compensates for the non-trivial measure induced by the AdS geometry.

The SM-like term (\ref{3.2}) is not the sole contribution to the masses 
of the zero modes. The KK states of the fermions couple to the
zero modes and induce corrections to (\ref{3.2}) via tree-level and loop
diagrams. 
We require these additional contributions to be small compared to (\ref{3.2}).
The tree-level contribution,
arising due to a mixing of the zero modes and the excited fermion
states, is of order \cite{dAS} 
\begin{equation} \label{tree}
\delta m_f^{(0)}\sim\lambda_{0i}  \lambda_{ij}  \lambda_{0j}\frac{v^3}{M_iM_j},
\end{equation}
where $M_{i,j}$ are the masses of the excited fermions, $v=174$ GeV is the
4d Higgs vev and $\lambda_{ij}$ are the Yukawa-type couplings between 
the fermion modes and the Higgs in the 4d effective theory.  
The corrections to fermion masses from a loop involving a Higgs 
and KK fermion states has been estimated in \cite{DHR2},
\begin{equation} \label{loop}
\delta m_f^{(1)}\sim \frac{1}{16\pi^2}\lambda_{0i}  \lambda_{ij}  \lambda_{0j}v.
\end{equation}

Since the KK fermions are  localized closer to the
TeV-brane than the zero modes, their couplings to the Higgs boson
is enhanced. Therefore, it is not clear that the fermion mass corrections
involving these states are small, even though these states 
are rather heavy. Requiring that the corrections (\ref{tree}) and (\ref{loop})
are small compared to the fermion mass (\ref{3.1}) provides additional 
constraints on the c-parameter of the fermions. 

For the parameter sets (A) and (B) of table \ref{t_L} we find that the 
additional contributions are indeed small, $\delta m_f^{(0)}/m_f\sim0.03$ and 
$\delta m_f^{(1)}/m_f\sim0.08$, irrespective of the lepton flavor. In this case 
the fermion masses can be reliably calculated from eq.~(\ref{3.2}).

If we increase the 5d Yukawa coupling, the fermions can be moved
closer towards the Planck-brane. However, at the same time the fermion
mass corrections (\ref{tree}) and (\ref{loop}) increase exponentially. 
We find that fine-tuning of the fermion mass contributions is only 
avoided for $\lambda^{(5)}/g^{(5)}\lsim5$. For the muon, 
for instance, this means that $c_L<0.614$ $(c_L<0.707)$ for the left-right
symmetric $(c_E=1/2)$ case. In this parameter range the constraints
arising from the anomalous muon magnetic moment are satisfied as well 
\cite{DHR2}.   

In the case of quarks we have to take into account the additional
complication of flavor mixing.\footnote{For sake of simplicity we ignore 
CP-violation here. It is readily included by
introducing complex Yukawa couplings for the quarks.} 
The mixing provides additional constraints on the 5d quark mass
parameters. There are nine $c$-parameters in the quark sector, three each
for the left-handed doublets $c_Q$, the right-handed $u$-quarks $c_U$,
and the right-handed $d$-quarks $c_D$. The physical quantities we want
to reproduce are the six quark masses and, in the absence of CP-violation,  
the three CKM mixing angles. 

Rather then keeping the Yukawa couplings strictly equal to one, as in the case
of leptons, we allow them to vary by a factor of two, 
i.e.~$1/2<|\lambda^{(5)}_{ij}|/g^{(5)}<2$, which certainly introduces no
new hierarchies. One parameter set which reproduces the physical quark masses
and mixings \cite{pdg} is given by
\begin{eqnarray}
c_{Q1}=0.72 & c_{D1}=0.57 & c_{U1}=0.63  \nonumber \\
c_{Q2}=0.60 & c_{D2}=0.57 & c_{U2}=0.30  \nonumber \\
c_{Q3}=0.35 & c_{D3}=0.60 & c_{U3}=0.10,  \label{ps}
\end{eqnarray}
\begin{eqnarray}
\frac{\lambda^{(5)}_D}{g^{(5)}}=\left(\begin{array}{ccc} 
0.50 & -2.00 & -2.00 \\
1.48 & -0.90 & 2.00 \\
0.52 & -0.50 & 0.70
\end{array}\right), \quad
\frac{\lambda^{(5)}_U}{g^(5)}=\left(\begin{array}{ccc} 
0.80 & -1.90 & -2.00\\ 
1.23 & 1.20 & -1.04 \\
1.85 & 1.66 & -0.80 
\end{array}\right). \nonumber
\end{eqnarray}
Using these numbers we obtain
\begin{eqnarray}
&m_u=2.9~{\rm MeV}, \quad & m_c=1.3~{\rm GeV}, \nonumber \\
&m_d=3.8~{\rm MeV}, \quad & m_b=4.4~{\rm GeV}, \nonumber \\
&m_s=78~{\rm MeV}, \quad & m_t=165~{\rm GeV}, \nonumber 
\end{eqnarray} 
\begin{equation}
|V_{\rm CKM}|=\left(\begin{array}{ccc} 
0.9744 & 0.2248 & 0.0045 \\
0.2248 & 0.9736 & 0.0392 \\
0.0045 & 0.0392 & 0.9992
\end{array}\right).
\end{equation}
We have checked that for this parameter set the additional mass
contributions from eqs.~(\ref{tree}) and (\ref{loop}) are sub-leading. 
Like in the case of leptons an overall enhancement of the 5d
Yukawa couplings by a factor of about five is tolerable. For larger 
Yukawa couplings, however,  (\ref{tree}) and (\ref{loop}) become 
important.

Strictly speaking the quark masses given above are running
masses at the cutoff scale of the effective 4d theory, which is
in the TeV range. However, the effects of evolving the 
masses to the low energy regime could be absorbed 
in a redefinition of the 5d Yukawa mass matrices.  
The localization of the right-handed top-quark at the TeV-brane 
is shown in figure \ref{f_1}.

Note that the parameter set  (\ref{ps}) is not uniquely determined
by the experimental constraints. But it demonstrates that 
bulk fermions in the RS model can naturally explain with order
one parameters not only the huge fermion mass hierarchy but
also the quark mixings.  
Note that in the RS model a profile for fermions  
in the extra dimension is automatically induced by the 
non-factorizable geometry. As a result, this scenario is
quite constrained, as will become clear in the 
discussion of the impact on non-renormalizable operators in the 
next section.

\section{Dimension six operators and proton decay}
We consider the following generic four-fermion operators 
which are relevant for proton decay and $K-\bar K$ mixing
\begin{equation}
\int d^4x \int dy \sqrt{-g}\frac{1}{M_5^3}\bar \Psi_i\Psi_j\bar\Psi_k\Psi_l
\equiv \int d^4x \frac{1}{M_4^2}\bar \Psi_i^{(0)}\Psi_j^{(0)}\bar\Psi_k^{(0)}\Psi_l^{(0)}
\end{equation}
where the effective 4d mass scale is given by
\begin{equation}
\frac{1}{M_4^2}=\frac{1}{2\pi^2kR^2M_5^3}\frac{1}{N_0(c_i)N_0(c_j)N_0(c_k)N_0(c_l)}
\frac{e^{(4-c_i-c_j-c_k-c_l)\pi kR}-1}{4-c_i-c_j-c_k-c_l}.
\end{equation}
Let us first consider 4-fermion operators built from a single lepton
flavor. 
For the left-right symmetric case and $M_5=k$ the results for $M_4$ are given
in  the last line of table \ref{t_L}. One observes the rough relationship $M_4 \propto 1/m$.
Furthermore we have $M_4 \propto \lambda^{(5)}/g^{(5)}$ and $M_4 \propto M_5^{3/2}$. 

$M_4$ also depends on the width of the Higgs profile. The further
the Higgs profile stretches out to the Planck brane (the smaller $a$ gets), 
the closer the fermions can be moved to the Planck-brane, and
the larger the suppression scale becomes. However, $a$ has to be
considerably smaller than four in order to change the given results.
For instance, with $a=2$, the effective $M_4$ is increased only
by a factor of 2.5. Raising the suppression scale of a 4-fermion operator involving
only electrons, for example,
to $10^{16}$ GeV requires $c=0.92$, which can be achieved only for $a\leq1.4$.
In this case the Higgs profile compensates the warp factor to a large extent. 
However, it is unclear if such a profile can be derived from a more 
fundamental scheme.  

Let us now discuss some specific operators. Constraints from 
$K-\bar K$ mixing require that the dimension-six operator $(d\bar s)^2/M_4^2$
is suppressed by $M_4\gsim 10^6$ GeV. Using the $c$-parameters of
({\ref{ps})  we obtain $M_4=5.5\times10^6$ GeV, in agreement 
with the constraint. This conclusion was also reached
in ref.~\cite{GP}. However, our estimate for the suppression scale
is more realistic, since we have taken into account the quark mixings.

Concerning proton decay the most stringent constraints arise from 
the  four-fermion operators
$O_L$=$Q_1Q_1Q_2L_3$ $(M_4>10^{15})$ GeV and 
$O_R$=$U_1^cU_2^cD_1^cE_3^c$ $(M_4>10^{12})$ GeV \cite{BD}.
Using the $c$-parameters of (\ref{ps}) and case (B) of table \ref{t_L} 
we obtain for these operators the effective suppression scales $M_4(O_L)=7.7\times10^8$ GeV
and $M_4(O_L)=1.7\times10^6$ GeV, which
are several orders of magnitude smaller than the experimental limits. If we
enlarge all the  5d Yukawa coupling by a factor of 5, the 
fermions can be moved further towards the Planck brane while still generating
their desired masses. $M_4$ then increases by a factor of 5.  Taking
$M_5=10k$ instead of $M_5=k$ would enlarge $M_4$ by another factor
of about 30. If the two factors are combined, the effective suppression scales 
can be pushed to $M_4(O_L)=1.2\times10^{11}$ GeV
and $M_4(O_L)=2.6\times10^8$ GeV, which is still four orders of 
magnitude below the constraint.  Larger values of $\lambda^{(5)}$ and
$M_5$ would further increase this result, but at the price of introducing
new hierarchies.  The same holds for a larger variation
in the 5d Yukawa couplings. Moreover, for $\lambda^{(5)}/g^{(5)}\gsim5$
the KK fermion contributions to the fermion masses (\ref{tree})
and (\ref{loop}) can no longer be ignored.

We conclude that by letting the SM fermions reside in the extra dimension
considerable suppression  of the non-renormalizable operators responsible
for proton decay can be achieved. However, the effective suppression scale of 
these operators is still too small by at least four orders of magnitude.
A small number of order $10^{-8}$ has to multiply the dimension-6 operators
to satisfy the experimental constraints on the proton life time. 

\section{Dimension five neutrino masses}
Majorana masses for left-handed neutrinos are generated by the 
dimension-five operator
\begin{equation} \label{nu3}
\int d^4x\int dy \sqrt{-g}\frac{l_{ij}}{M_5^2}H^2\Psi_{iL}C\Psi_{jL}
\equiv \int d^4x ~m_{\nu ij}\Psi_{iL}^{(0)}C\Psi_{jL}^{(0)},
\end{equation}
where $l_{ij}$ are some dimensionless couplings constants, 
$C$ is the charge conjugation operator and
\begin{equation} \label{nu}
m_{\nu ij}=\int_{-\pi R}^{\pi R}\frac{dy}{2\pi R}\frac{l_{ij}}{M_5^2}e^{-4\sigma(y)}
H^2(y)f_{0i}(y) f_{0j}(y).
\end{equation}
Because of the SU(2) symmetry, the 5d Dirac mass parameters
of the left-handed neutrinos and charged leptons are equal. 
No new parameters enter the game. For the left-right symmetric
case (A) of table \ref{t_L},  taking $M_5=k$  and $l_{ij}=\delta_{ij}$, 
i.e.~ignoring neutrino mixing, we obtain 
\begin{equation}
m_{\nu_e}=39 {\rm ~keV}, \quad m_{\nu_{\mu}}=8.8 {\rm ~MeV}, \quad  m_{\nu_{\tau}}=150 {\rm ~MeV}.
\end{equation}
These neutrino masses are well above the experimental limits.

The situation improves 
if the left-right symmetry is given up, and left-handed fields
are moved closer to the Planck-brane.  
Let us consider case (B) where the right-handed fields
are not localized at all $(c_E=0.5)$. Then 
the 5d mass parameters of the left-handed leptons read
$c_e=0.834$, $c_{\mu}=0.664$ and $c_{\tau}=0.567$.
We then obtain the neutrino masses
\begin{equation}\label{nu2}
m_{\nu_e}=2.3 {\rm ~eV}, \quad m_{\nu_{\mu}}=112 {\rm ~keV}, \quad  m_{\nu_{\tau}}=33 {\rm ~MeV}.
\end{equation}
While these neutrino masses may be compatible with
collider experiments, they are problematic for neutrino
oscillations and  cosmology. 

From eq.~\ref{nu} one observes that $m_{\nu} \propto 1/M_5^2$.
For $c_R=0.5$ we also have  $m_{\nu} \propto 1/(\lambda^{(5)})^2$. 
Assuming $M_5=10k$ (instead of $M_5=k$) and $\lambda^{(5)}/g^{(5)}=5$,
we could reduce the neutrino masses by another factor 
of $2.5\times 10^3$. However, the $\nu_{\tau}$ mass is still in the keV region, and
another factor of $10^5$ is needed to bring it down to the 
range suggested by the atmospheric neutrino oscillation data. However, with the
required large numbers, $M_5=100k$,  $\lambda^{(5)}/g^{(5)}=100$,
new hierarchies arise in the model, and large corrections to fermions masses 
arise from eqs. (\ref{tree}) and (\ref{loop}). We conclude that some symmetry
(e.g. lepton number) is needed to prevent large
dimension five neutrino masses in the Randall-Sundrum model.

Finally, let us assume that with suitably  large values of $M_5$ and $\lambda^{(5)}$ the 
neutrino masses of eq.~\ref{nu2} are indeed reduced by a factor of
$10^8$. Can the operator of eq.~\ref{nu3}
explain the neutrino oscillation data?  
Including non-diagonal terms in (\ref{nu3}) by setting all $l_{ij}=1$ we obtain
the following mass matrix in the $(\nu_e,\nu_{\mu},\nu_{\tau})$ basis
\begin{equation}
m_{\nu}= \left(\begin{array}{ccc} 
2.28 & 505 & 8.66~10^3 \\
505 & 1.12~10^5 & 1.92~10^6 \\
8.66~10^3 & 1.92~10^6 & 3.31~10^7
\end{array}\right) 10^{-8}~{\rm eV.}
\end{equation}
As one might expect from the geometrical picture, the neutrino mass matrix
is of nearest neighbor type, similar to those of the quarks.
The corresponding mixings are in the few percent range, in conflict
with the atmospheric neutrino data. If we relax the
constraint on the couplings in  (\ref{nu3}) to $1/2 <|l_{ij}|<2$ the
situation improves to some extent. The $\nu_e$--$\nu_{\mu}$
$(\nu_{\mu}$--$\nu_{\tau})$
mixing angle can become as large as $\pi/5$ $(\pi/15)$.
But these values are still too low to explain the data. 
Moreover, we were not able to find a parameter 
set where both mixings are large at the same time, which seems 
favored by the data. In order to explain
the experimental results the field content of the model most likely 
has to be extended. The right-handed neutrino
is an obvious choice (see, e.g.~\cite{GN}), and we will explore 
this in a future publication.
   
\section*{Acknowledgement}
S.~H.~is supported in part by the Feodor Lynen Program of the
Alexander von Humboldt foundation. This work was also supported 
by DOE under contract DE-FG02-91ER40626.

\end{document}